\documentclass[twocolumn,showpacs,prl]{revtex4}
\usepackage{graphicx}

\usepackage{bm}
\usepackage{amsmath}
\usepackage{amsfonts}
\newcommand{\ch}{{\cal H}}
\newcommand{\ck}{{\cal K}}

\newcommand{\tr}{{\rm Tr}}
\newcommand{\ket}[1]{| #1 \rangle}
\newcommand{\bra}[1]{\langle #1 |}
\newcommand{\braket}[2]{\langle #1 | #2 \rangle}

\begin{document}
\title{Simple security proof of 
quantum key distribution via uncertainty principle}

\author{Masato Koashi}
\affiliation{Division of Materials Physics, Graduate School
of Engineering Science, Osaka University, 1-3 Machikaneyama, 
Toyonaka, Osaka 560-8531, Japan}
\affiliation{CREST Photonic Quantum Information Project, 4-1-8 Honmachi, Kawaguchi, Saitama 331-0012, Japan}

\begin{abstract}
We present an approach to the unconditional security of quantum key
distribution protocols based on the uncertainty principle.
The approach applies to every case that has been treated 
via the argument by Shor and Preskill, and relieve them from the
constraints of finding quantum error correcting codes. It can 
also treat the cases with uncharacterized apparatuses. We derive 
a secure key rate for the Bennett-Brassard-1984 protocol with 
an arbitrary source characterized only by a single parameter 
representing the basis dependence.

\pacs{03.67.Dd 03.67.-a}
\end{abstract}
\maketitle

The aim of quantum key distribution (QKD) is to distribute 
a secret key between two distant parties, Alice and Bob, 
under the intervention by a third party, Eve.
For any protocol of QKD, it is vital to have a proof of the
unconditional security because the robustness against 
any kind of attack
allowed by the law of physics is the main advantage of QKD over
classical schemes aiming at the same task. 
One of the well-known strategies for the security proof is 
the argument \cite{Shor-Preskill00}
 given by Shor and Preskill, in which a reduction to an entanglement
distillation protocol (EDP) based on Calderbank-Shor-Steane (CSS)
quantum error correcting codes (QECC)
\cite{Calderbank-Shor96,Steane96_2551}
 is used to show that the
information leak on the final key is negligible.
This approach has turned out to be quite versatile
due to the simplicity of the idea: for example,
the original proof for the BB84 protocol \cite{Bennett-Brassard84}
has been extended \cite{TKI03,Koashi04}
to cover the B92 protocol \cite{Bennett92}.
On the other hand, 
invoking the CSS-QECC in the proof requires the actual users
 to find 
a quantum code satisfying a certain property, which is not always an easy 
task. Even the innocent-looking formula
[Eq.~(\ref{eq:SPkeygain}) below]
 for the asymptotic key gain
needs a complicated argument \cite{Hamada03} for strict derivation.

If we look back to the first proof \cite{Mayers96}
 of unconditional security by
Mayers, we notice that it also has its own merits. One disadvantage, the
complexity of the proof, was recently remedied by a simple 
proof \cite{Koashi-Preskill03} by
Koashi and Preskill based on the same spirit, namely, reduction to a
two-party protocol by omitting one of the legitimate users by a symmetry
argument. In this line of approach, the error correction and the privacy
amplification is decoupled once we encrypt the communication for the
error correction, by consuming the previously shared key. This implies
that we do not need to find a CSS-QECC and we can just use conventional
schemes for the error correction. The proof also shows a peculiar
and useful property, which allows the use of basis-independent uncharacterized
sources or detectors. For example, if we use an ideal detector, the
source can be anything as long as it does not reveal which basis is used
in the BB84 protocol. We can still use the same formula for the key
rate, indicating that any fault in the source can be automatically
caught in the form of an increase in the observed bit errors. Unfortunately, the
argument of omitting one party relies heavily on the symmetry of the
BB84 protocol, and it cannot be applied to the protocols with no such
symmetry. 

In this paper, we present an approach to the unconditional security
based on uncertainty principle. This argument has the same
advantages in the Mayers-Koashi-Preskill argument, while retaining the
versatility of the Shor-Preskill argument. In fact, 
in any protocol having a proof that relies on the Shor-Preskill
argument, we can decouple the error correction and the privacy
amplification just by encrypting the former, thereby relieve it from the
constraint of CSS-QECC. We can also treat uncharacterized apparatuses in
the protocols with lower symmetry. 
As an example, we derive a key rate
formula for the BB84 protocol with an arbitrary source, 
the properties of which are unknown except for 
a bound on a single parameter describing the basis dependence.

Most of the QKD protocols can be equivalently 
described by an entanglement-based protocol, 
in which 
Alice and 
Bob share a pair of quantum systems 
$\ch_A\otimes \ch_B$ after discarding 
other systems used for random sampling tests. 
The state $\rho_0$ of 
$\ch_A\otimes \ch_B$ at this point is 
not fixed 
and may be highly correlated
among subsystems
due to Eve's intervention, but 
the results of the tests may give 
a set of promises on the possible state.
For example, in the case of Shor-Preskill 
proof, $\ch_A\otimes \ch_B$ is composed of 
$N$ pairs of shared qubits, and there
is a promise that the following statements
hold except for an exponentially small probability:
Suppose that each qubit 
is measured on $z$ or $x$ basis. Then the number 
$n_{\rm bit}$ of qubits showing the bit error
($\sigma_z \otimes \sigma_z=-1$) satisfies
$n_{\rm bit}/N \le \delta_{\rm bit}$,
and   
the number $n_{\rm ph}$ with
the phase error 
($\sigma_x \otimes \sigma_x=-1$)
satisfies
$n_{\rm ph}/N \le \delta_{\rm ph}$.
Here $\delta_{\rm bit}$ and $\delta_{\rm ph}$
are determined from the results of the test.
Here we consider more general cases, in which 
the size of $\ch_A\otimes \ch_B$ is arbitrary. 
We give a proof for the unconditional security
of the protocols having the following form:

{\it Actual Protocol} ---
Alice and Bob make measurements on $\ch_A$ and on $\ch_B$,
respectively. Through an encrypted classical 
communication consuming $r$ bits of secret key, they 
agree on an $N$-bit reconciled key $\bm{\kappa}_{\rm rec}$, 
except for a negligible failure probability.
In the binary vector space on $N$ bits, one 
party chooses a linearly-independent 
set $\{\bm{V}_k\}_{k=1,\ldots, N-m}$ of $N$-bit sequences
randomly and announce it.
The $k$-th bit of the final key $\bm{\kappa}_{\rm fin}$ 
is defined as scalar product
$\bm{\kappa}_{\rm rec}\cdot \bm{V}_k$.

This protocol newly produces $N-m$ bits of secret key,
and the net secret key gain is $N-r-m$ bits.
The core of our approach is to 
choose a quantum operation $\Lambda$ that converts 
state $\rho$ on $\ch_A\otimes \ch_B$ to state
$\Lambda(\rho)$ on $\ch_R\otimes\ck^{\otimes N}$, 
where $\ck^{\otimes N}$ stands for $N$ qubits and $\ch_R$
for an ancillary system $R$. Both the qubits and the ancilla 
are virtual, and there is no need to specify corresponding
physical systems in the actual protocol. We allow $\Lambda$
to involve collective operations over $\ch_A$ and $\ch_B$.
We only require the following property for $\Lambda$:
Let us regard $\bm{\kappa}_{\rm rec}$ in Actual Protocol
 as the outcome 
of a generalized measurement applied on 
$\ch_A\otimes \ch_B$. Then, the application of $\Lambda$
followed by the $z$-basis
measurements on the $N$ qubits should be equivalent to this measurement
of $\bm{\kappa}_{\rm rec}$.  
If $\Lambda$ is chosen in this way, the security of Actual Protocol
follows that of Protocol 1 below, in which Alice 
and Bob can be regarded as a single party:

{\it Protocol 1} --- 
Apply $\Lambda$ and discard $\ch_R$. For the $N$
qubits $\ck^{\otimes N}$,
measure each qubit on $z$-basis to determine
the $N$-bit key $\bm{\kappa}_{\rm rec}$.
Choose a linearly-independent 
set $\{\bm{V}_k\}_{k=1,\ldots N-m}$
randomly, and announce it to Eve.
Let $\bm{\kappa}_{\rm rec}\cdot \bm{V}_k$
be the $k$-th bit of the final key $\bm{\kappa}_{\rm fin}$.

In order to show that Eve has negligible information on 
$\bm{\kappa}_{\rm fin}$, we consider yet another protocol.
Suppose that, following $\Lambda$,
 we conduct a
measurement $M_R$ on the ancilla $R$ to obtain outcome 
$\mu$, and subsequently measure each of the $N$ qubits on $x$ basis
 to obtain an $N$-bit sequence $\bm{X}$.
We further choose a number $\xi$
(depending on the test results)
 such that 
 the promise on the initial state
$\rho_0$ almost guarantees that for each outcome $\mu$,
we can predict the value of $\bm{X}$ with $N\xi$-bit 
uncertainty. More precisely, we take the following 
assumption:

{\it Assumption} --- There exists a set $T_\mu$ of $N$-bit 
sequences with cardinality $|T_\mu|\le 2^{N\xi}$ for each $\mu$, such 
that the pair of measurement outcomes $(\mu, \bm{X})$ satisfies
$\bm{X}\in T_\mu$ except for an exponentially small probability 
$\eta$.

Now we can invoke the uncertainty principle: 
Since the $x$-basis outcomes for the $N$ qubits can be 
predicted with $N\xi$-bit uncertainty, the complementary 
observable, namely, the $z$-basis outcomes, should be 
predicted by any party with at most $N(1-\xi)$
uncertainty \cite{Maassen-Uffink88}. 
Hence we expect to extract $N(1-\xi)$ bits 
of secret key from the $z$-basis outcomes. This rough
sketch can be made strict as follows.

Suppose that, before the measurement of $\bm{X}$,
we choose $m= N(\xi+\epsilon)$ 
random $N$-bit sequences $\bm{W}_j (j=1,\ldots,m)$
and measure the parity $\bm{X}\cdot \bm{W}_j$ by a collective 
projection measurement on the qubits. If we define $\Sigma_\nu(\bm{W})
\equiv \sigma_\nu^{b_1}\sigma_\nu^{b_2}\cdots\sigma_\nu^{b_N}
(\nu=x,z)$ for 
$N$-bit sequence $\bm{W}=[b_1b_2\cdots b_N]$, the above 
parity measurement for $\bm{X}\cdot \bm{W}_j$
 corresponds to the observable $\Sigma_x(\bm{W}_j)$.
Recall that we know $\bm{X}\in T_\mu$ except for probability $\eta$.
As in the hushing method of EDP \cite{BDSW96},
by knowing $m$ random parity bits we can 
derive an estimate $\bm{X}^*$ of $\bm{X}$
with an exponentially 
small failure probability 
$Pr(\bm{X}^*\neq \bm{X})\le 
\eta'\equiv \eta+2^{-N\epsilon}$.
If we apply a phase-flip operation 
$\Sigma_z(\bm{\bm{X}^*})$ according to the
estimate, the state $\sigma$ of the qubits
should become almost a pure state, satisfying
$\bra{0_x^{\otimes N}}\sigma \ket{0_x^{\otimes N}}
\ge 1-\eta'$, where $\ket{0_x^{\otimes N}}$
is the $x$-basis eigenstate for $\bm{X}=\bm{0}$.
With this property in mind, let us consider the following 
protocol:

{\it Protocol 2} --- 
Apply $\Lambda$ and make measurement $M_R$ on $\ch_R$.
Choose $\bm{W}_j (j=1,\ldots m)$ randomly,
and take an arbitrary linearly-independent 
set $\{\bm{V}_k\}_{k=1,\ldots N-m}$ of $N$-bit sequences
satisfying $\bm{V}_k\cdot \bm{W}_j=0$ for any $j,k$.
Announce $\{\bm{V}_k\}$ to Eve.
Measure $\Sigma_x(\bm{W}_j)$ to determine $\bm{X}^*$,
and apply $\Sigma_z(\bm{\bm{X}^*})$. 
Measure $\{\Sigma_z(\bm{V}_k)\}$ to determine 
the $(N-m)$-bit final key $\bm{\kappa}_{\rm fin}$.

When Assumption holds with $\epsilon >0$,
the above final key is determined by $z$-basis measurements 
applied to $\sigma$, which is very close to the
$x$-basis pure eigenstate $\ket{0_x^{\otimes N}}$. 
Hence Eve has only negligible 
(at most $S(\sigma)$-bit)
information about $\bm{\kappa}_{\rm fin}$.

The equivalence of the two protocols are easy to be seen.
In Protocol 2, the operators $\{\Sigma_z(\bm{V}_k)\}$ commute
with $\Sigma_z(\bm{\bm{X}^*})$ and with $\Sigma_x(\bm{W}_j)$
since $\bm{V}_k\cdot \bm{W}_j=0$. 
Hence we can omit the parity check and the phase flip and still 
obtain the same final key. We further notice that 
$M_R$ is now redundant, and the choosing method of $\{\bm{V}_k\}$ 
can be simplified to a random selection. 
Noting that $\{\Sigma_z(\bm{V}_k)\}$ can be 
also obtained through a $z$-basis measurement on each qubit,
we are lead to Protocol 1. We thus obtain the main theorem:

{\it Theorem} --- If Assumption is true for $m=N(\xi+\epsilon)$
with $\epsilon>0$, Eve's information on $\bm{\kappa}_{\rm fin}$
in Protocol 1 is at most $h(\eta')+N\eta'$ with 
$\eta'=\eta+2^{-N\epsilon}$.

Here we have defined $h(y)\equiv -y\log y - (1-y)\log (1-y)$.
The choice of $\Lambda$ and $M_R$, which determines $\xi$, is crucial 
in deriving a good lower bound of the achievable secure 
key gain for various problems. We will discuss several 
examples below.

{\it Shor-Preskill case} --- In the situation to which 
the Shor-Preskill argument applies, $\ch_A\otimes \ch_B$
corresponds to $N$ pairs of qubits.  In this case, 
we choose  $\bm{\kappa}_{\rm rec}$ to be Bob's measurement 
outcome on $z$ basis.
If the promise is given by the two numbers $\delta_{\rm bit}$
and $\delta_{\rm ph}$ as we mentioned earlier, Alice 
can determine $\bm{\kappa}_{\rm rec}$ from her $z$-basis
measurement and $r=N[h(\delta_{\rm ph})+\epsilon]$ bits of 
communication from Bob in Actual protocol. For the security 
proof, we choose 
 a trivial $\Lambda$ that just changes the 
definition as $\ch_A\cong \ch_R$ and $\ch_B\cong \ck^{\otimes N}$.
We assume $M_R$ to be
the $x$-basis measurement on Alice's $N$ qubits.
It should reveal the value of $\bm{X}$, which is 
Bob's outcome on $x$ basis,
within $\delta_{\rm ph}$ bits of errors, and Assumption 
holds with $\xi=h(\delta_{\rm ph})+\epsilon$.
Hence we arrive at the familiar asymptotic net key gain 
\begin{equation}
G=N[1-h(\delta_{\rm bit})-h(\delta_{\rm ph})].\label{eq:SPkeygain}
\end{equation}
Unlike the Shor-Preskill proof, this key rate is achieved
without finding a CSS-QECC. 

{\it BB84 with a basis-independent uncharacterized source}
--- This is the case where Alice uses a basis-independent
uncharacterized source and Bob uses an ideal detector 
in the BB84 protocol, which was analyzed in \cite{Koashi-Preskill03}.
Let $\rho_{ab}$ acting on $\ch_Q$ be the state of Alice's source for the 
basis $a=0,1$ and the bit value $b=0,1$. 
Alice chooses basis $a$ randomly, and then 
with probability $p_{ab}$ (note that $p_{a0}+p_{a1}=1$), Alice 
sends out $\rho_{ab}$ to a quantum 
channel, which may be tampered by Eve. Bob 
receives a qubit state on $\ck_B$ from the channel, on which he 
conducts the ideal $z$- or $x$-basis measurement depending 
on his random basis choice $a'=0,1$, respectively.
Subsequently, they make $a$ and $a'$ public. 
After repeating this many times, 
they randomly sample events with $a=a'$ to determine
the observed error rates $\delta_a$ for $a=0,1$. 
Bob randomly picks $N$ outcomes from the unsampled 
data with $a=a'=0$ to define $\bm{\kappa}_{\rm rec}$. 
Alice obtains $\bm{\kappa}_{\rm rec}$ with the help of 
a secret communication from Bob consuming
$r=N[h(\delta_{0})+\epsilon]$ bits of secret key.
(The portion with $a=a'=1$ can be handled similarly.)

The basis-independent 
source satisfies $\rho_{0}=\rho_{1}$,
where $\rho_a\equiv p_{a0}\rho_{a0}+p_{a1}\rho_{a1}$.
Then, we can find a state $\chi$ on 
$\ch_S\otimes \ch_Q$ and measurements $M_a$ on 
$\ch_S$ with POVM elements $\{F_{a0},F_{a1}\}$,
such that $\tr_S[(F_{ab}\otimes \bm{1}_Q)\chi]=p_{ab}\rho_{ab}$.
We are thus allowed to consider an equivalent protocol 
in which Alice prepares $\chi$ and conducts  measurement $M_a$
on $\ch_S$ to determine her bit value $b$. 
This new protocol takes the form of Actual protocol by 
defining $\ch_A=\ch_S^{\otimes N}$ and $\ch_B=\ck_B^{\otimes N}$.
 For the security 
proof, we choose 
 a trivial $\Lambda$ that just changes the 
definition as $\ch_A\cong \ch_R$ and $\ch_B\cong \ck^{\otimes N}$.
We then assume $M_R$ to be $M_1$ applied on each $\ch_S$.
In order to establish a statement like Assumption,
we need to know the relation between the outcome of $M_1$ and 
the outcome of the $x$-basis measurement on $\ck_B$. Fortunately, this 
is exactly the same pair of measurements used in determining 
 the error rate $\delta_1$. Hence Assumption holds with
$\xi=h(\delta_{1})+\epsilon$, and 
we obtain the asymptotic net key gain
\begin{equation}
G=N[1-h(\delta_{0})-h(\delta_{\rm 1})].
\end{equation}
Note that everything we need in the actual protocol is 
$\delta_0$ and $\delta_1$. There is no need to
know the identities of $\chi$ and $M_a$, and hence
no need to characterize the source to determine $\rho_{ab}$,
as long as it is guaranteed to be basis-independent.

{\it BB84 with a basis-dependent uncharacterized source}
--- The main theorem allows us to prove unconditional 
security in the general case of $\rho_0\neq \rho_1$. 
Of course, we need to know something about the source states
since the protocol is entirely insecure if $\rho_0$ and $\rho_1$
are orthogonal. A natural choice is to assume that we know 
a single parameter $\Delta$, which determines
 a lower bound on the fidelity
\cite{Uhlmann76,Jozsa94} between the two states:
\begin{equation}
1-2\Delta\le \sqrt{F(\rho_0,\rho_1)}\equiv \tr (
\sqrt{\rho_1}\rho_0 \sqrt{\rho_1}
)^{1/2}.
\end{equation}
Note that for $F< 1$, 
we can still find two pure states $\ket{\chi_0}$ and 
$\ket{\chi_1}$ in 
$\ch_S\otimes \ch_Q$ satisfying $\braket{\chi_0}{\chi_1}=1-2\Delta$
such that for each value of $a$, 
there is a POVM measurement $M_a=\{F_{a0},F_{a1}\}$ on 
$\ch_S$
satisfying $\tr_S[(F_{ab}\otimes \bm{1}_Q)\ket{\chi_a}\bra{\chi_a}]
=p_{ab}\rho_{ab}$.
For a special case where 
 $\ch_S$ includes a qubit as a subsystem and 
$M_0$ and $M_1$ are the standard $x$- and $z$-basis
measurement on that qubit, 
Gottesman {\it et al.} \cite{GLLP02}
derived a secure key rate along the line of Shor-Preskill argument,
which allows positive key gain 
up to $\Delta<0.029$. 
Here we can derive a better 
key rate formula for arbitrary states $\{\rho_{ab}\}$.

Let us consider an equivalent protocol in which 
Alice chooses the basis $a$ by measuring a 
``quantum coin'' \cite{GLLP02} described by
a qubit $\ck_C$. If she prepares 
$\ch_S\otimes \ch_Q\otimes \ck_C$ in state
$\ket{\Psi}\equiv 
(\ket{\chi_0}\ket{0_z}_C+
\ket{\chi_1}\ket{1_z}_C)/\sqrt{2}$ and  
 measure $\ck_C$ on $z$ basis, the outcome 
$a$ is random and $\ch_S\otimes \ch_Q$ is
prepared in state $\ket{\chi_a}$.
Then she conducts measurement $M_a$ on $\ch_S$ to prepare
$\rho_{ab}$ with probability $p_{ab}$.
In order to prove security, we follow the same 
argument as in the basis-independent case
up to the point where we need to know the relation
between the outcome of $M_1$ and that of 
$x$-basis measurement on $\ck_B$.
Unfortunately, we have no direct clue this time.  
The expected error rate $\delta_{\rm ph}$ in 
this fictitious set of measurements is 
no longer equal to $\delta_1$, since the former 
is taken for $a=0$ and the latter is for $a=1$. 

In order to determine upper bounds on $\delta_{\rm ph}$,
let us consider the following scenario.
Alice starts from $\ket{\Psi}^{\otimes L}$, 
and she immediately sends the $L$ copies of system $Q$ 
into the channel. After Eve's attack, 
Bob receives the qubits $\ck_B^{\otimes L}$.
For every pair of systems $\ch_S\otimes \ck_B$,
Bob may choose $a'$ randomly, but regardless of 
its value, measurement $M_1$ and $x$-basis measurement 
are applied to determine whether there is an error 
($t=1$) or not ($t=0$). Finally, Alice measures 
the coin $\ck_C$ on $z$ basis to determine $a$.
Let us denote the empirical probability for 
the $L$ events by $r(\cdot)$. For example, 
$r(t=1|a=0)$ is the number of events with 
$(t=1,a=0)$ divided by that of events with $a=0$.

The rate $\delta_{\rm ph}$ can be regarded as 
an error rate in a fair sampling from the 
events with $a=a'=0$. Since $a'$ has no effect 
in the above scenario, it can also be regarded 
as a fair sampling from the events with $a=0$.
We thus have $\delta_{\rm ph}\cong r(t=1|a=0)$.
Similarly, $\delta_{1}\cong r(t=1|a=1)$.
Since $r(a=0)\cong 1/2$, we have
\begin{multline}
 r(t=1)\cong (\delta_{1}+\delta_{\rm ph})/2, \;\;
r(a=1|t=1)\cong\delta_1/(\delta_{1}+\delta_{\rm ph}), 
\\
r(a=0|t=0)\cong (1-\delta_{\rm ph})/(2-\delta_{1}-\delta_{\rm ph}).
\end{multline}

Now we describe two methods of deriving
a bound on $\delta_{\rm ph}$. The first one 
is to apply the main theorem formally to the 
coins, regarding $\ck_C^{\otimes L}$ as $\ck^{\otimes N}$
in the theorem. 
Since $\|{}_C\bra{1_x}\ket{\Psi}\|^2=\Delta$,
it is guaranteed that we can distill a secret key of 
length $L(1-h(\Delta)-\epsilon)$ from the $z$-basis measurement 
results. This implies that even with the knowledge of 
each $t$, the entropy of the outcomes $a$ should be larger than 
$L(1-h(\Delta)-\epsilon)$. Hence we have 
\begin{multline}
 1-h(\Delta)\le \frac{\delta_{1}+\delta_{\rm ph}}{2}h\left(
\frac{\delta_{1}}{\delta_{1}+\delta_{\rm ph}}
\right)
\\
+
\frac{2-\delta_{1}-\delta_{\rm ph}}{2}h\left(
\frac{1-\delta_{\rm ph}}{2-\delta_{1}-\delta_{\rm ph}}
\right)
\le 
h\left(
\frac{1-|\delta_{\rm ph}-\delta_1|}{2}
\right),
\end{multline}
which shows that $\delta_{\rm ph}=\delta_1$ for 
$\Delta=0$ and $\delta_{\rm ph}$ becomes larger 
when $\Delta>0$. If we write the maximum of 
$\delta_{\rm ph}$ under the above first inequality 
as $f(\delta_1, \Delta)$, the 
key gain is given by 
\begin{equation}
G=N[1-h(\delta_{0})-h(\max\{1/2,f(\delta_1, \Delta)\})].\label{eq:keyrate}
\end{equation}
This key gain is positive only for $\Delta<0.056$.

The second method is more complicated, but gives a better rate.
We assume that for each event, 
Alice draws a random binary variable $s$ with  
a small probability of being $s=1$. 
If $s=0$, she just follows the above scenario, but
if $s=1$, she measures the coin $\ck_C$ on $x$ basis instead of 
$z$ basis. Let $\bar{a}$ be the outcome of this $x$ 
basis measurement, and define 
$r_{x,j}\equiv r(\bar{a}=1|s=1,t=j)$ and 
$r_{z,j}\equiv r(a=0|s=0,t=j)$ for $j=0,1$.
Since $\|{}_C\bra{1_x}\ket{\Psi}\|^2=\Delta$,
we have
\begin{equation}
 r(t=0)r_{x,0}+r(t=1)r_{x,1}=r(\bar{a}=1|s=1)\cong \Delta.\label{eq:delta}
\end{equation}
Note that $r_{x,j}$ is determined from the 
outcomes of $x$-basis measurements applied to
random samples from the qubits with $t=j$, and
$r_{z,j}$ is from the $z$-basis outcomes 
for the rest of the qubits. This problem of 
random sampling was analysed in \cite{TKI03},
and it was shown that for all $\epsilon>0$,
except for an exponentially 
small probability, there exists a qubit state
$\rho$ such that 
$|r_{z,j}-\bra{0_z}\rho\ket{0_z}|< \epsilon$
and $|r_{x,j}-\bra{1_x}\rho\ket{1_x}|< \epsilon$.
We thus obtain the following relation 
in the asymptotic limit:
\begin{equation}
 (1-2r_{x,j})^2+(1-2r_{z,j})^2\le 1.
\end{equation}
Combining it with 
$r(t=1)\cong (\delta_1+\delta_{\rm ph})/2$, 
$r_{z,1}\cong \delta_{\rm ph}/(\delta_{1}+\delta_{\rm ph})$,
$r_{z,0}\cong (1-\delta_{\rm ph})/(2-\delta_{1}-\delta_{\rm ph})$,
and Eq.~(\ref{eq:delta}), we obtain
\begin{equation}
 2\Delta\ge 1-\sqrt{(1-\delta_{1})(1-\delta_{\rm ph})}
-\sqrt{\delta_{1}\delta_{\rm ph}}.\label{eq:2delta}
\end{equation}
We can now take $f(\delta_1,\Delta)$ to be the 
maximum of $\delta_{\rm ph}$ under Eq.~(\ref{eq:2delta}),
and obtain a better key rate with Eq.~(\ref{eq:keyrate}).
Now the region of positive key gain extends to 
$\Delta<0.146$, or $F(\rho_0,\rho_1)>1/2$.
Since Alice and Bob do not use the outcome $\bar{a}$,
this measurement can be omitted. Hence, in the actual 
BB84 protocol, they only have to discard a small portion of 
events. From Eve's point of view, Alice {\it could have} 
measured $\bar{a}$ for the discarded events, and it is enough 
to apply the above security proof.

We have described a method of proving the unconditional 
security which unifies two major previous approaches 
and retains the advantages in both of them. 
We have also shown that
the new method can solve a problem which eluded the previous 
approaches. 
The proof relies on the observation that Alice can guess
the $z$-basis outcomes of virtual $N$ qubits with $r$-bit
uncertainty in the actual protocol, and Alice and Bob 
can guess the $x$-basis outcomes with $m$-bit uncertainty
in a equivalent protocol. The ``excess'' over the 
uncertainty limit, $N-r-m$, amounts to the key gain.
Note that if they share a maximally entangled state, 
Alice alone can guess for both of the bases. The condition 
for the secrecy is weaker than that since it allows 
her to collaborate with Bob nonlocally for the $x$ basis.
This difference is considered to be a reason for the gap
between distillable entanglement and secret 
key gain \cite{HHHJ05}. This suggests that the present 
method may potentially give a key rate exceeding the amount
of distillable entanglement.

The author thanks N.~Imoto and J.~Preskill for helpful 
discussions. This work was supported by a MEXT Grant-in-Aid 
for Young Scientists (B) 17740265.

\bibliographystyle{apsrev}

\end{document}